\documentclass[a4paper,10pt]{article}

\usepackage{textcomp}
\usepackage{graphicx}
\usepackage{amsmath}
\usepackage{amsfonts}
\usepackage{amssymb}

\title{\textbf{High-Temperature Optical Constants of Dust Analogues for the Solar Nebula}}

\date{}

\begin{document}

\maketitle

\begin{flushleft}

\textit{Manuscript status:} submitted as contribution for the ECLA proceedings (European Conference on Laboratory Astrophysics from 26.-30. September 2011 in Paris)  in the \textit{EAS Publication Series}
\newline
\newline
\textit{Authors:} Simon Zeidler$^1$, Harald Mutschke$^1$
\newline
$^1$Astrophysikalisches Institut, Schillerg\"asschen 2-3, D-07745 Jena, Germany\\
(simon.zeidler@uni-jena.de, harald.mutschke@uni-jena.de)

\abstract{The dust in protoplanetary disks is influenced by a lot of different processes. Besides others, heating processes are the most important ones: they change not only the physical and chemical properties of dust particles, but also their emission spectra. In order to compare observed infrared spectra of young stellar systems with laboratory data of hot (up to 700\textdegree C) circumstellar dust analogues, we investigate materials, which are important constituents of dust in protoplanetary disks. We calculated the optical constants by means of a simple Lorentzian oscillator fit and apply them to simulations of small-particle emission spectra in order to compare our results with real astronomical spectra of AGB-stars and protoplanetary disks.}

\end{flushleft}

\section{Background}

Matter in space is subjected to various different temperature conditions. The temperature has a strong effect on the absorption/emission behavior of dust particles \cite{Koike} and this effect could thoroughly help to give explanations of still unidentified bands in stellar spectra (for instance the 13\,$\mu$m band in AGB-star spectra). Band shifts may also indicate the emission originating from different parts of circumstellar disks. Therefore, studying the influence of the temperature on  the  absorption properties of astrophysical relevant solid materials is an important aim of laboratory astrophysics.

\section{Performing high temperature infrared spectroscopic measurements}

\begin{figure}
 \begin{center}
 \includegraphics[width=0.8\linewidth]{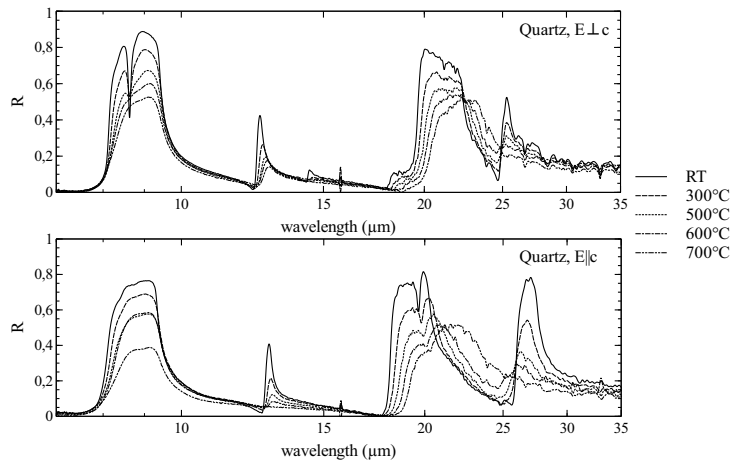}
 \end{center}
\caption{The measured reflectivity spectra of the natural $\alpha$-quartz crystal in both polarizations. At 700\textdegree C some bands totally disappeared (e.g. the 27\,$\mu$m band in E$\parallel$c polaritation), which resulted from the second order phase transition to $\beta$-quartz.}
\label{quartz}
\end{figure} 

In this project, which is part of the DFG SPP "The first 10 million years of the solar system", we investigated the dependence of the optical constants on the temperature for synthetic bulk materials such as Corundum ($\alpha$-Al$_2$O$_3$) and Spinel (MgAl$_2$O$_4$) as well as for natural minerals ($\alpha$-Quartz (SiO$_2$, from Brazil) and Olivine (Mg$_{1.84}$Fe$_{0.16}$SiO$_4$, San Carlos) crystals). Corundum and spinel prevail as early condensates in the stellar outflow of AGB-stars and are constituents of the CAIs in meteorites \cite{Fabian01}. Olivine is a very important constituent of circumstellar dust disks \cite{Juhasz}\cite{Henn} around young and evolved stars and of our own planetary system (main constituents of meteorites \cite{Sunshine}).

All samples had oriented (except spinel) and polished surfaces to perform infrared reflection-spectroscopy with polarized light. In order to inplement these measurements while the samples are heated, we used a special High-Temperature-High-Pressure (HTHP) cell, that is built into our Bruker 113v FTIR Spectrometer. The sample chamber of the cell is a cylindric hole in the sample heater of 13\,mm diameter through that the infrared beam can reach the sample. An infrared polarizer has been set right behind the cell in the outcoming beam, which realized the reflection measurements under polarized light.

The measurements have been taken at only four different temperatures: room temperature (RT), 300\textdegree C, 500\textdegree C, and 700\textdegree C to prevent the heater from erosion. Only in the case of $\alpha$-quartz we measured also at 600\textdegree C to get a better temperature resolution of the (second order) phase transition to $\beta$-quartz that takes place between 500 and 700\textdegree C (see Figure \ref{quartz}).

\section{Derivation of optical constants}

\begin{figure}
 \begin{center}
 \includegraphics[width=0.8\linewidth]{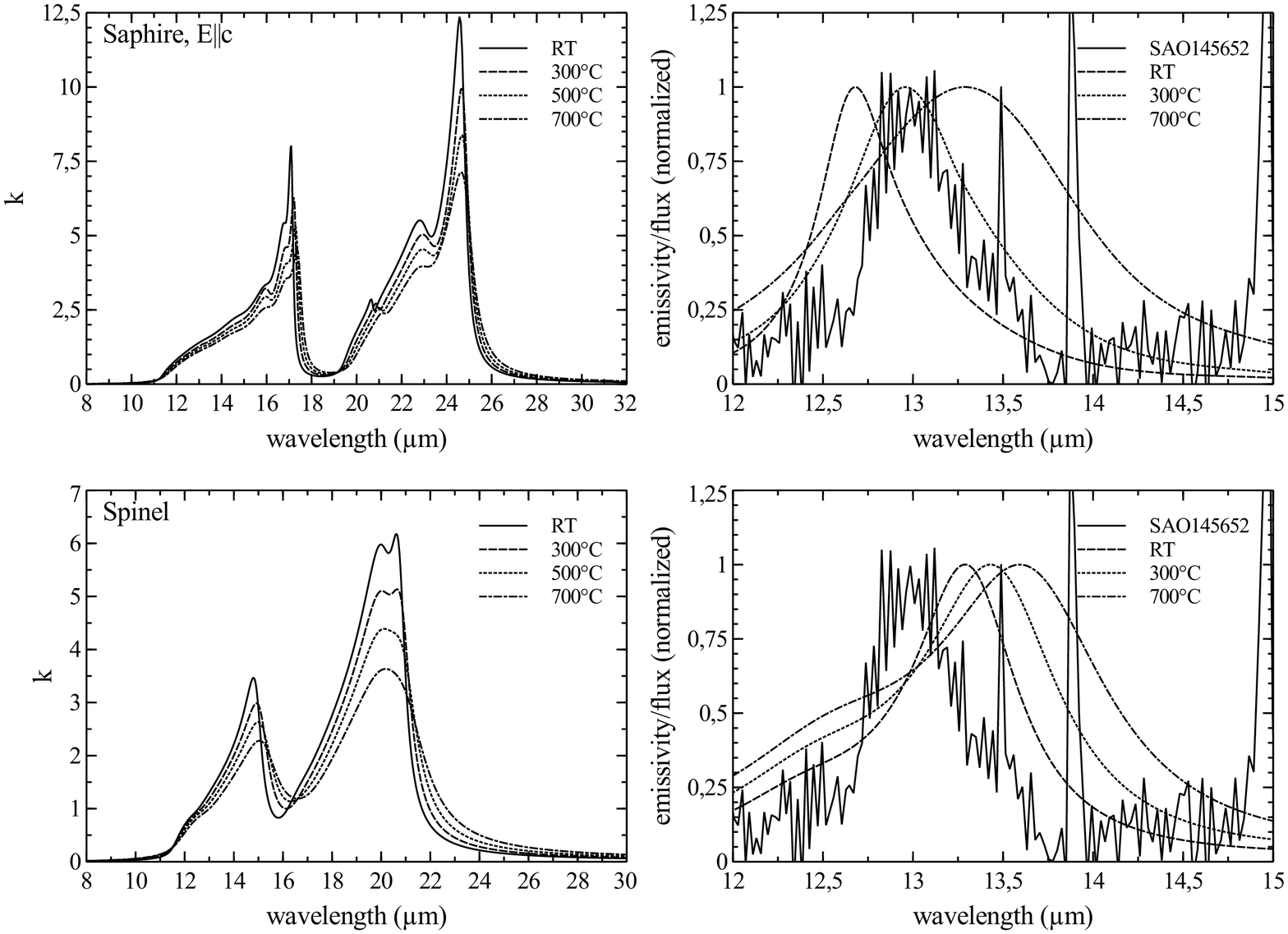}
 \end{center}
\caption{Top: on the left side, the absorption cross section (Cabs) for Corundum (top) and Spinel (below) from room temperature (RT) up to 700\textdegree C can be seen. On the right side, the comparison between the normalized emission spectra of the 13\,$\mu$m band of the AGB star SAO145652 and the normalized emission of Corundum (top) and Spinel (below) is shown for RT, 300, and 700\textdegree C (Cabs multiplied with the corresponding Planck-Function).}
\label{cor_spi}
\end{figure} 

From the measurements we optained simple reflection spectra as can be seen in Figure \ref{quartz}. The optical constants (the real part $n$ and the imaginary part $k$ of the complex refractive index) have been derived from these spectra with Lorentzian oscillator fits. 

Furthermore, the temperature dependence of the fitting parameters itself could be approached by fits of second order polynomials, which resulted in a full analytical description of the temperature dependent optical constants. We also calculated the absorption cross section $C_{abs}(T)$ for small ($a\ll\lambda$) spherical particles and multiplied $C_{abs}(T)$ with the Planck-function at each temperature in order to get the emissivity spectra and to compare the results with astronomical spectra.

The results for Spinel and Corundum, can be seen in Figure \ref{cor_spi}. The $k$-spectra for Corundum and Spinel show almost the same behavior with increasing temperature. The bands shift to longer wavelengths and the intensity of the bands decreases. They also become slightly broader. We also compared the small particle emissivity spectra of both materials with the emission spectrum of the AGB star SAO145652, focusing on the 13\,$\mu$m band (see Figure \ref{cor_spi} on the right side). The best reproduction of the 13µm band with respect to the band position is only reached by Corundum at 300\textdegree C. At this temperature the band of Corundum is too broad to fit the 13\,$\mu$m band of the star perfectly, but this effect could either be a reason of the calculations or have its origin in the noise of the spectrum of the star. 

\section{Application to the mineralogy of a protoplanetary disk}

Besides pyroxenes, olivines are the most abundant crystalline minerals in circumstellar dust disks \cite{Juhasz}\cite{Henn}. However, the exact composition of these olivines is still under discussion (especially the iron content), since the composition itself, the grain shape, and the grain size have a strong influence on the position and the shape of bands. With our investigations we can show, that also  the temperature, as an additional parameter, plays an important role for the analysis of spectra of protoplanetary disks and should be attended in future disussions about the composition of circumstellar dust. 

\begin{figure}
 \begin{center}
 \includegraphics[width=0.7\linewidth]{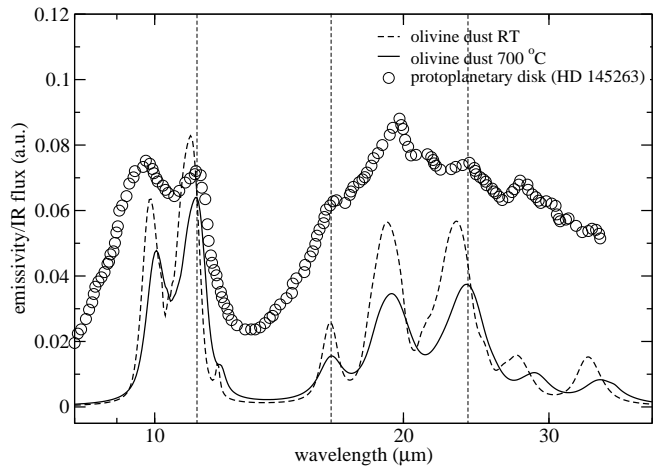}
\end{center}
\caption{Comparison of calculated small particle emissivity spectra of olivine at two different temperatures and the emission spectrum of the protoplanetary disk HD145263.}
\label{olivine}
\end{figure}

In Figure (\ref{olivine}) we compared the emission spectrum of the protoplanetary disk of HD145263 (taken with ISO) with calculated small particle emissivity spectra of olivine, based on our measurements at RT and 700\textdegree C. It can be seen that the dust at 700\textdegree C fits somewhat better the observed emission spectrum than the dust at RT (indicated by the vertical dashed lines, which mark the band positions of olivine dust in the spectrum of HD145263). Of course we do not claim, that the olivine dust in this protoplanetary disk has a temperature of exactly 700\textdegree C. Beside the other mentioned influences on an emission spectrum, the dust will always hold a certain temperature distribution. But it should be noted, that for the analysis of spectra of circumstellar dust disks, the temperature of the dust should be taken into account just as the shape, size, and the composition of the dust grains.

\end{document}